\documentclass{article}
\usepackage{arxiv}

\usepackage[utf8]{inputenc} 
\usepackage[T1]{fontenc}    
\usepackage{hyperref}       
\usepackage{url}            
\usepackage{booktabs}       
\usepackage{amsfonts}       
\usepackage{nicefrac}       
\usepackage{microtype}      
\usepackage{color}
\usepackage{graphicx}
\usepackage{float}
\usepackage[font=small]{caption}
\usepackage{amsmath}
\usepackage{amssymb}
\usepackage{mathtools}
\usepackage{tabularx}
\usepackage[round, semicolon, authoryear]{natbib}
\let\rho\varrho 

\title{A dynamic network model of societal complexity and resilience inspired by Tainter’s theory of collapse}

\author{
  Florian Schunck
  \thanks{Helmholtz Centre for Environmental Research GmbH---UFZ, Research Group System Ecotox, Permoserstraße 15, 04318 Leipzig; Osnabrück University, Institute of Mathematics, Research Group System Science, Barbarastraße 12, 49076 Osnabrück
  \texttt{florian.schunck@uni-osnabrueck.de}},  
  Marc Wiedermann
  \thanks{Potsdam Institute for Climate Impact Research, Member of the Leibniz Association, FutureLab on Game Theory and Networks of Interacting Agents, P.O. Box 60 12 03, D-14412 Potsdam, Germany}, 
  Jobst Heitzig
  \thanks{Potsdam Institute for Climate Impact Research, Member of the Leibniz Association, FutureLab on Game Theory and Networks of Interacting Agents, P.O. Box 60 12 03, D-14412 Potsdam, Germany},  
  Jonathan F. Donges
  \thanks{Potsdam Institute for Climate Impact Research, Member of the Leibniz Association, FutureLab on Earth Resilience in the Anthropocene, P.O. Box 60 12 03, D-14412 Potsdam, Germany; Stockholm Resilience Centre, Stockholm University, Kr\"aftriket 2B, 106 91 Stockholm, Sweden
  \texttt{donges@pik-potsdam.de}}
 }

\begin{document}
\maketitle

\begin{abstract}
    In recent years, several global events have severely disrupted economies and social structures, undermining confidence in the resilience of modern societies. Examples include the COVID-19 pandemic, which brought unprecedented health challenges and economic disruptions, and the emergence of geopolitical tensions and conflicts that have further strained international relations and economic stability. While empirical evidence on the dynamics and drivers of past societal collapse is mounting, a process-based understanding of these dynamics is still in its infancy. Here we aim to identify and illustrate the underlying drivers of such societal instability or even collapse. The inspiration for this work is Joseph Tainter's theory of the ``collapse of complex societies'', which postulates that the complexity of societies increases as they solve problems, leading to diminishing returns on complexity investments, and ultimately to collapse. In this work we have abstracted this theory into a low-dimensional and stylised model of two classes of networked agents, hereafter referred to as ``laborers'' and ``administrators''. We numerically modeled the dynamics of societal complexity, measured as the fraction of ``administrators'', which is assumed to affect the productivity of connected energy-producing ``laborers''. We show that collapse becomes increasingly likely as the complexity of the model society continuously increases in response to external stresses that emulate Tainter's abstract notion of problems that societies must solve. We also provide an analytical approximation of the system's dominant dynamics, which matches well with the numerical experiments, and use it to study the influence on network link density, social mobility and productivity. Our work advances the understanding of social-ecological collapse and illustrates its potentially direct link to an ever-increasing societal complexity in response to external shocks or stresses via a self-reinforcing feedback.
\end{abstract}



\section{Introduction}

Human societies have always faced a wide variety of challenges that have tested their resilience and resulted in a median longevity of premodern societies of around 200 years~\citep{Scheffer.2023}. These challenges include external stresses such as invasions and environmental catastrophes as in the case of Mesopotamia \citep{Weiss.1993} or internal pressures such as corruption, rebellion and mismanagement, as in ancient Egypt \citep{Butzer.2012}. Historically, both cases led to societal collapse. Other classic and widely studied examples include the fall of the Western Roman Empire in the 5th century CE, the collapse of the Maya in the 9th century CE or the fall of the Minoan civilization of Crete in the 14th century BCE \citep{Tainter.1988, Middleton.2012}. Many more historical civilizations and their trajectories have also been studied extensively in order to identify common factors of collapse \citep{Turchin.2003}.

Modern societies face similarly severe challenges, such as global climate change, pandemics, or financial instability. The estimated costs of climate change \citep{Tol.2018} alone will put additional strain on already over-indebted nation states \citep{WorldBank.2020}. While the 2008 housing crisis still reverberated in several countries, the Covid-19 pandemic has caused major impacts on health systems and may have long-term consequences for economies, political institutions and social structures \citep{Trump.2020}.

The above examples illustrate the long-standing question of whether it is possible to identify underlying principles that determine a society's ability to cope with such large-scale challenges and thus increase its resilience to collapse. As reviewed by \citet{Cumming.2017}, there are several explanatory models for collapse. Explanations focusing on resource limitations such as the classical Malthusian trap \citep{Malthus.1798} and limits to exponential growth and overuse of resources have been widely discussed in the past \citep{Meadows.1972, Rockstrom.2009, Diamond.2011, dixson2022earth}. But climate change, inequality, and other hypotheses also point to collapse \citep[e.g.][]{Kennett.2012, Acemoglu.2015}. These theories can be broadly divided into external and internal drivers of collapse \citep{Scheffer.2023}.

Tainter's {\em theory of the collapse of complex societies} specifically emphasizes the notion of societal complexity and its tendency to self-amplify in response to stress as the primary cause of collapse \citep{Tainter.1988, Tainter.2006}. Within this framework, complexity emerges continuously through problem solving and is manifested in the differentiation and specialization of social roles, hierarchies and control of behavior, growing population, technical abilities and increased information flows. 
These investments in complexity have associated costs, such as calories, natural resources, time, or money, all of which can be reduced to an abstract form of \textit{energy}. Although increases in complexity can be beneficial to societies and contribute to well-being, it is hypothesized that the surpluses generated by increases in complexity will diminish as the ``low-hanging fruits'' become fewer. Thus, diminishing marginal returns to investments in complexity (or simply returns on complexity $ROC$) ultimately drive a society into collapse due to loss of resilience to external perturbations or internal crises. In addition, the benefits of complexity may be short-lived and quickly consumed by population growth or increased living standards. In contrast, increased energy demand tends to be persistent~\citep{Acemoglu.2008}, leading to new problems once previously earned energy surpluses are used up. Within the framework proposed in~\citet{Tainter.1988}, this process is described as the {\em energy-complexity spiral} \citep{Tainter.2006}.

Contemporary evidence for complexity gains in specialization and differentiation of social roles as well as technical abilities are the decline of labor in the agricultural sector and increase of the financial and communications sectors in the 20th century, while  productivity in primary sectors increased \citep{Feinstein.1999}. Further evidence of complexity increase is the growing burden of bureaucracy. In the U.S., the number of administrative regulations increased linearly from 1950--2000 by approximately 2500 pages per year \citep{Coglianese.2002}. For example, in cancer research, complexity is introduced by formalities or regulations for opening trials. Such complexity increments diminish the returns on investment (ROI) of research, resulting in reduced innovativeness in the field \citep{Steensma.2014}. \citet{Hall.2009} estimates the energy return on investment (EROI) of a sustainable society to be well above 3:1, which can be interpreted as a proxy for complexity. Despite these examples, overall empirical evidence on societal complexity and its change over time is scarce, even though it seems ubiquitous in our lives.

Mathematical models and simulations are important tools for exploring the implications of different theories of collapse and resilience and for testing competing hypotheses about the processes underlying the dynamics of societal complexity and collapse. In the archaeological domain, agent-based models are employed to understand the interaction of water and land-use change \citep{Barton.2010} or to compare trajectories of biocultural evolution \citep{Barton.2012}. Dynamical systems models are frequently employed to model small-scale social-ecological systems \citep{Kohler.2007, Ostrom.2009, Schlueter.2012}. Examples include a resource economics model of the collapse of the Rapa Nui population \citep{Brander.1998} or an exploration of vulnerabilities in forager-resource systems \citep{Freeman.2012}. More recently, \citet{Motesharrei.2014} have developed a dynamical systems model to study trajectories to collapse in a coupled hierarchical consumer-resource system. Similarly, stylized models of coupled carbon, population, and capital stocks have been used on a planetary scale to identify preconditions for societal collapse and sustainability and their implications for World-Earth system resilience~\citep{Nitzbon.2017,anderies2023modeling}.

In the last decade, a new generation of models has been conceptualized turning to larger and up to global scales and embracing social-ecological complexity by integrating concepts from dynamic land-use models and agent-based models \citep{Rounsevell.2011, Arneth.2014}. Building on this framework \citet{Brown.2019} identified societal breakdown as an emergent property of large-scale behavioral models of land-use change under different climate-economic scenarios. To strengthen the human side of the equation(s), much attention has been paid to increase realism in modeling human decision making in socio-economic and social-ecological models \citep{Schluter.2017, Schill.2019,beckage2022incorporating,moore2022determinants} as well as modeling collective behavior and social tipping dynamics \citep{Wiedermann.2020,otto2020social}. Recently, these efforts have culminated in a comprehensive modeling framework for so-called World-Earth Models \citep{Donges.2020,donges2021taxonomies}. Also, Tainter's theory was previously modeled following a system dynamics based approach, focusing on the diminishing returns of resource acquisition, due to the formation of a bureaucracy \citep{Bardi.2019}. 

In this work, we take an alternative approach and explore the theory of {\em collapse of complex societies} in terms of the mechanisms that lead to increasing societal complexity, diminishing resilience and ultimately collapse. To this end, we implement an agent-based network model consisting of productive and administrative nodes. This implementation successfully recaptures the theory's postulated behavior of collapse due to diminishing marginal returns to complexity. We then study its macroscopic dynamics in detail and identify two escape mechanisms from collapse: increasing the output elasticity of labor or allowing social mobility by stochastic transitions of nodes between productive and administrative states. In order to approach the theory of societal collapse from a conceptual angle, we propose here to reduce it to three key elements: 
\begin{enumerate}
    \item The basic currency of any society is energy, since labor and material goods can be viewed as driven by or derived from energy.
    \item Problems need to be solved when energy availability is deficient as a consequence of stochastic events or shocks. According to the self-reinforcing process denoted as \textit{energy-complexity spiral} (see above) this process always increases societal complexity.
    \item Increases in complexity can be modeled as increases in administrative capacity because they encapsulate increases in specialization of social roles, hierarchies and control, and information flow. Moreover, the size of an administrative body relative to the overall size of the system is a very tangible example of complexity.
\end{enumerate} 

The remainder of this paper is organized as follows. We begin with a thorough description of our proposed network model in the next section. We first present the basic model and then describe a model variant that includes a mechanism to counteract collapse. In the results section, we first present the stochastic model by discussing exemplary trajectories of the system followed by the derivation of an analytical approximation. Based on this approximation we present a comprehensive ensemble analysis of all crucial model parameters. The paper concludes with a discussion of all relevant results and an outlook on future work.


\section{Model description} 

\subsection{Tainter-inspired network model of a hierarchical society steering into diminishing mar-ginal returns and collapse} 

\begin{figure}[!t]
    \centering
    \includegraphics[width=.65\linewidth]{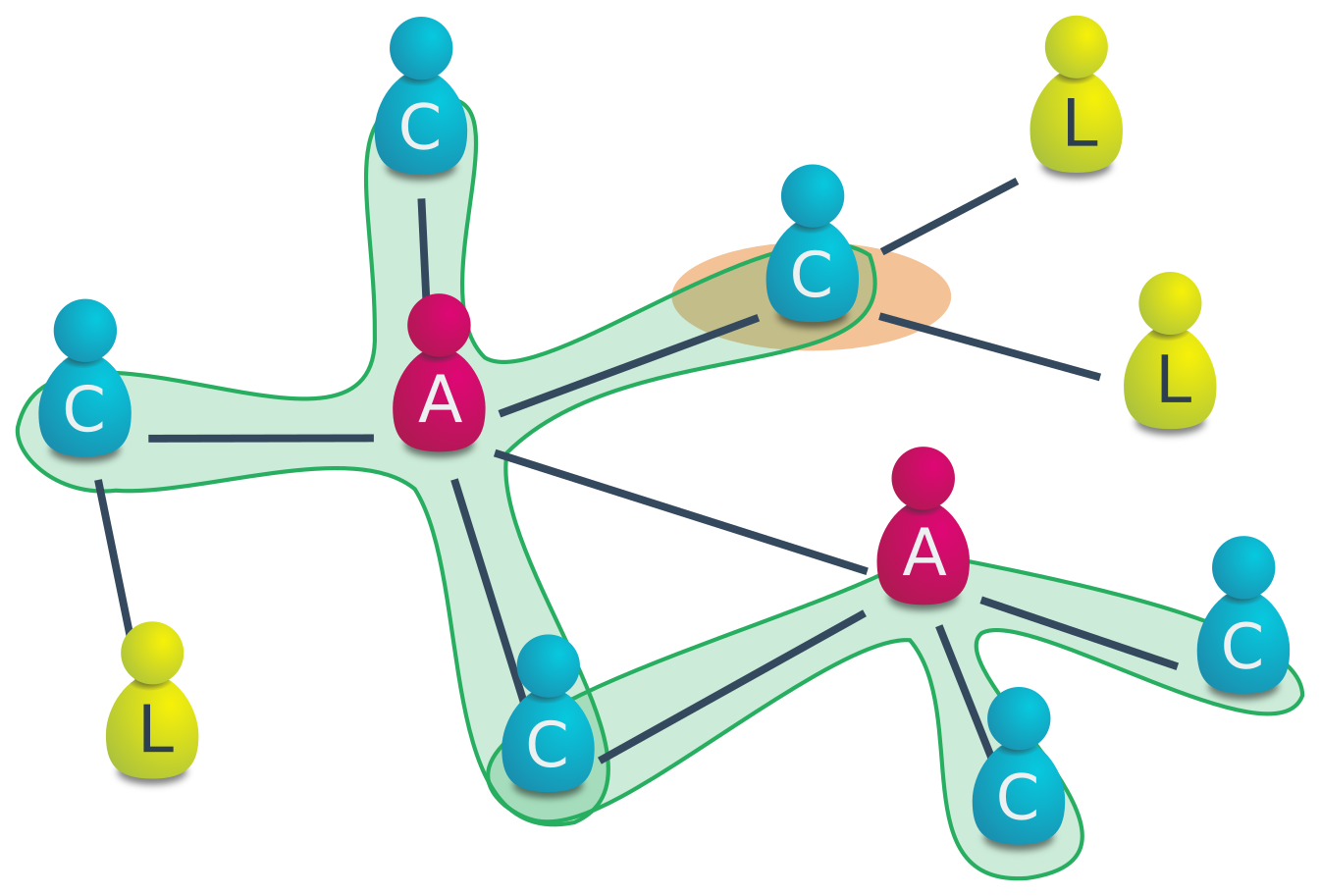}
    \caption{Sketch of the model setup. Nodes/agents are either attributed as \textit{administrators} $A$ or \textit{laborers} $L$. laborers that are connected to at least one administrator are denoted as \textit{coordinated laborers} $C$ that produce higher energy outputs due to an increased productivity. The influence of each administrator is marked with green shading. In case the energy production falls below a critical value, the coordinated laborer that is connected with most other nodes (here marked in orange) becomes an administrator as well.}
    \label{fig:example_network}
\end{figure}

In the following we introduce the model that is used in this work to exemplify the dynamics of collapse that are proposed in~\citet{Tainter.1988}. Its goal is to illustrate the increase in societal complexity as a response to the solving of problems, here the sustaining of
a certain level of energy supply $E$. We define complexity as an increase in the diversification of the society (the creation of a new role) and the corresponding increase in control over the behaviour of others (an increasing administrative body). 
We will show that with only including these two processes, one already observes an increase in an appropriately defined measure of complexity, and, hence, we do not need to specifically account for additional phenomena, such as specialisation or an increase in hierarchies.

In particular, we represent an artificial society by means of a complex network~\citep{newman_structure_2003, castellano_statistical_2009} as complex networks have been successfully used in the past to model social systems with respect to collapse~\citep{heckbert_mayasim:_2013}, sustainability~\citep{wiedermann_macroscopic_2015, barfuss_esd_2017, holstein2020optimization} and their ability to adapt to new conditions~\citep{auer_dynamics_2015, schleussner_clustered_2016}. They are usually comprised of entities of two basic types, i.e., nodes and links. For our case the nodes indicate (representative) small but long-lived entities that supply labor, such as family lines, in a society. Links between the nodes or entities represent some
comparatively stable
social tie, e.g., a professional relationship, between them. $N$ counts the total number of nodes in the network, Fig.~\ref{fig:example_network}.

The sole goal of the society is to produce energy $E$ at a certain level in order to sustain its function. The corresponding per-`capita' (actually, per-node) energy requirement of the society is denoted as $\epsilon$. In order to produce this energy certain nodes (denoted as laborers $L$) harvest an individual energy source $R_{max}$. $R_{max}$ is subject to external shocks that represent an abstract form of a problem that needs to be solved. In order to increase the energy that is produced from the resource, some laborers $L$ can also be appointed to become administrators $A$. The purpose of the administration is to increase the productivity of the laborers $L$ that are under its direct management.
We assume each laborer $L$ who is directly connected with at least one administrator $A$
becomes part of a more productive body of the labor force which has a larger total factor productivity and at least the same returns to scale, expressed by a productivity factor $c>1$ and an output elasticity of labor of $b$, where the rest of the labor force has an output elasticity of labor of $a\le b$. In line with common economic models, we assume overall decreasing returns to scale, i.e., $a\le b<1$.
We denote this class of more productive laborers the \textit{coordinated laborers} $C$, Fig.~\ref{fig:example_network}. While they increase the output elasticity of their immediate surrounding, the administrators $A$ themselves do not produce energy any longer (thus indicating a certain cost of maintaining an administrative body). Note that being additionally connected with more than one administrator $A$ does not further increase the productivity $c$ of the coordinated laborers $C$. Along the lines of ~\citet{Tainter.1988} we measure the resulting societal complexity $S$ in terms of the size of the administration $N_A=|A|$ as it represents the level of behavioral control in our model society.

We model the temporal dynamics within our artificial society according to the following rules. At each time step $t$ the maximally available resource per node, $R_{max}$, is reduced by randomly drawn disturbances from a beta distribution with parameters $\alpha = 1$ and $\beta = 15$. 

\begin{align}
    R = R_{max} (1 - B), ~\mathrm{with}~B \sim \mathrm{Beta}(\alpha,~\beta) \label{eqn:resource}
\end{align}

The stochasticity of $R$ here represents the occurrence of shocks or problems to be solved by the model society. Larger values of $B$ represent larger shocks or problems. We have chosen the parameters for the beta distribution such that drawing shocks mostly results in $R$ close to $R_{max}$ (i.e., no or small problems) with a rare chance of $R \ll R_{max}$ (i.e., large problems). Depending on the number of laborers $N_L(t)$ (uncoordinated) and $N_C(t)$ (coordinated) the total produced energy $E(t)$ is computed as

\begin{align}
E = R (N_L^a + c N_C^b),\label{eqn:energy}
\end{align}

where $a$ and $b$ determine the output elasiticity to scale of the nodes $L$ and $C$, respectively and $c$ is the productivity factor of coordinated workers $C$. $R_{max}$ is scaled by the output elasticity of uncoordinated workers $R_{max} = N / N^a$ to initialize the unperturbed per capita energy production at $E_{cap} = 1$. If energy per capita falls below a minimum value representing the society's vital needs, i.e., if $E(t)/N < \epsilon$ for some parameter $\epsilon$, we assume the society tries to solve this problem by appointing exactly one (additional) administrator such that the output elasticity of some laborers is increased. The new administrator is selected as follows:
 \begin{itemize}
    \item[(a)] If currently no administrators exist in the society (i.e., if $N_A=0$) the laborer $L$ with the most network connections becomes the sole administrator.
    \item[(b)] If at least one administrator currently exists (i.e., if $N_A>0$), the coordinated laborer who has the most network connections becomes an additional administrator.
\end{itemize}
The model then proceeds to the next time step and the availability of $R$ as well as the produced energy $E(t)$ is again computed according to Eqn.~\eqref{eqn:resource} and  Eqn.~\eqref{eqn:energy}.

Following~\citep{Tainter.2006} we define the potential onset of societal collapse, i.e., the vicinity to a critical state, once the system approaches diminishing returns on complexity $ROC$. We operationalize $ROC$ into our model by computing the differences in energy $E$ and complexity $S$ after the recruitment of one additional administrator $A$ at time $t$, 

\begin{align}
    ROC(t) = \frac{E(t) - E(t-1)}{S(t) - S(t-1)}.
\end{align}

Note that $ROC(t)$ depends on the specific structure of the underlying social network, in particular on who the newly appointed administrator $A$ is connected to and how many laborers turn from $L$ to $C$ as a consequence. As outlined above we measure societal complexity $S$ simply in terms of the size of the administration $N_A$, i.e., $S(t)=N_A(t)$. Since the size of the administration is increased by one due to the recruitment of a single additional administrator at time $t$ we obtain $N_A(t) = N_A(t-1)+1$ such that $S(t) - S(t-1) = 1$. Therefore the return on complexity $ROC$ reduces to $ROC(t) = E(t) - E(t-1)$.

Once $ROC(t)$ becomes negative our artificial society is expected to begin its decline into collapse since the energy production $E$ becomes smaller and approaches zero with increasing time $t$ and a correspondingly ever increasing size of the administration $N_A$. In the following section we therefore directly display results for the energy production $E(t)$ over time and interpret a positive slope as increasing returns to complexity and a negative slope as decreasing returns to complexity. In the scope of this model we define collapse as the state where $E = 0$. While this is unrealistic in real-world scenario, it serves as a reasonable endpoint for the abstract model proposed here, since qualitative differences are not expected when using collapse signals between $ROC \leq 0$ and $E=0$.

For the initial model setup we consider an Erd\H{o}s-R\'enyi random network~\citep{erdos_evolution_1960}. It consists of $N$ ($N=400$ for our case) nodes that are all initially assumed to be laborers ($N_L(0)=N$, $N_A(0)=N_C(0)=0$). Additionally, we place a link between each pair of nodes with a fixed probability $\rho$ (which we will vary between $\rho=0$ and $\rho=0.1$ for our analyses). Thus, $\rho$ gives the expected density of links in the resulting network, while $N(N-1)/2$ is the maximum number of possible links. Starting from this setup, we simulate the system for a maximum of $t_{max}=10,000$ time steps following the model logic that is given above. In the case when all nodes becomes administrators (i.e., if $N_A(t)=N$) we stop the simulation even though $t_{max}$ may not have been reached. In that case no more energy is produced ($E=0$ due to the lack of laborers $L$) and we consider the society to be entirely collapsed.

The code of the entire model and the described analysis is publicly available under \hyperlink{https://github.com/flo-schu/tainter}{https://github.com/flo-schu/tainter} including instructions for downloading pre-simulated datasets. Table \ref{tab:parameters} gives the parameters used in the simulation unless otherwise given.

\begin{table}[H] 
\caption{Model parameters, their function in the model and their values used in the analysis unless otherwise mentioned. The values in squared parentheses express the range scanned for the parameter analysis shown in Fig. \ref{fig:f4_parameters} \label{tab:parameters}}
\newcolumntype{S}{>{\hsize=.5\hsize}X}
\newcolumntype{C}{>{\centering\arraybackslash}S}
\newcolumntype{L}{>{\hsize=1.7\hsize}X}
\begin{tabularx}{\textwidth}{CLC}
\toprule
\textbf{parameter}	& \textbf{function}	& \textbf{value}\\
\midrule
$N$		& Network size (number of nodes)   & 400\\
$t_{max}$ & Maximal runtime of the simulation (time-steps) & 10,000\\
$\alpha$& Parameter of shock regulating Beta-distribution & 1\\
$\beta$ & Parameter of shock regulating Beta-distribution & 15\\
$\epsilon$ & Energy-threshold for coordinating a new $A$ & 1.0\\
$a$     & Output elasticity to scale of uncoordinated workers $L$ & 0.75\\
$b$     & Output elasticity to scale of coordinated workers $C$ & 0.75\\
$c$     & Productivity of coordinated workers $C$ & 1.05 $[1.0,...,3.0]$\\
$\rho$  & Link probability between nodes   & 0.02 $[0.0, ..., 0.1]$\\
$p_{e}$  & Exploration probability between node states   & 0.00 $[0.0, ..., 0.2]$\\
\bottomrule
\end{tabularx}
\end{table}

\subsection{Social mobility as a possible countermeasure to collapse}

The basic model that we introduced in the previous section represents closely the assumptions put forward in~\citet{Tainter.1988}. Here a society may get caught in the energy-complexity spiral as the increased or even flat demand for energy increases its complexity (measured as the size of the administrative body $N_A(t)$) when being faced with ever new problems.

One obvious downside of the above model setup is that, once selected, an administrator $A$ does not change its role back to become a laborer $L$ (or $C$) again (rachet effect). This decision is motivated by findings that social mobility is predominantly stable or upward rather than downward within \citep{Stawarz.2018} and across generations~\citep[e.g.,][]{adermon_intergenerational_2018}. Additionally, it is only possible to become an administrator $A$ if one is selected from the set of coordinated laborers $C$, meaning that a spontaneous jump in hierarchy is not permitted within the logic of the model.

However, in order to increase social mobility, we also aim to investigate a version of our model that allows for the nodes to randomly change their state with a fixed small probability. Such dynamics can be interpreted as the loss or gain in socio-economic status within or across generations. Within the logic of our model such an additional process is implemented as follows. At the beginning of each time step $t$ every node changes its state with probability $p_e$ either from $A$ to $L$ or $C$ (depending on whether it is connected to another administrator), or from $L$ or $C$ to $A$. We expect that this allows to regulate the complexity of the society and possibly avoid its unlimited growth (as for this setting the complexity may also decrease as the problems/shocks become smaller again). Hence, such process should cause a sustainable ratio between the size of the administration $N_A$ and the body of laborers $N_L+N_C$. Note that if we set $p_e=0$, we again obtain the dynamics of the model setup that is outlined in the model description.

\section{Results}

\subsection{Example trajectories} 

\begin{figure}[!t]
    \centering
    \includegraphics[width=\linewidth]{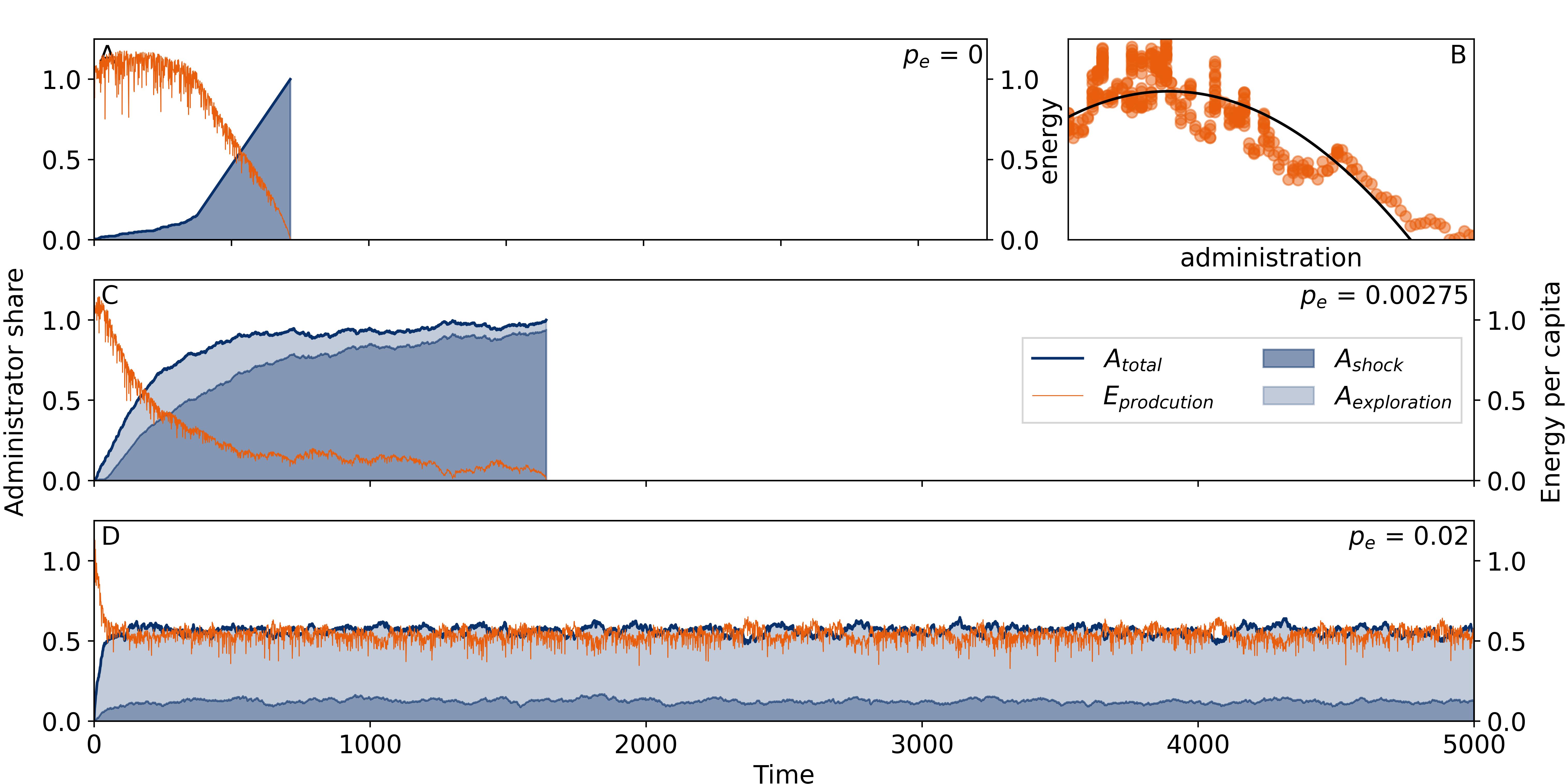}
    \caption{Exemplary network simulations of a Tainter-like model society ($N = 400$) according to the model description. The panels show only the first 5000 time steps to focus on the initial dynamic. Blue curves show the share of administration in the network (light blue: Administration as result of decreased resource availability, dark blue: Administration resulting from exploration with rate $p_e$). Orange curves show the average energy produced per node. A) shows the typical development of a network reacting to shocks by changing one node from coordinated laborer $C$ to administrator $A$. B) displays a moving average of the energy measured in A) against the size of the administration. The black curve indicates stylized parabola-shaped diminishing marginal returns as in ~\citep{Tainter.2006}. C) displays the network development at an intermediate exploration probability ($p_e = 0.00275)$. D) shows the development of a network high probability of exploration ($p_e = 0.02$).}
    \label{fig:fig2_exampleTrajectories}
\end{figure}

Figure \ref{fig:fig2_exampleTrajectories}A shows the implementation of the original approach to Tainter's theory of societal collapse with dynamics described above in the model description.
Initially the model society benefits from its response to external shocks. The reason is that initially each new administrator $A$ has a much higher number of connections to laborers $L$, which are thus turned into coordinated laborers $C$ producing energy output at increased productivity (here $a = 1.05$). As the amount of laborers $L$ decreases, the marginal returns on complexity ($ROC$) of each new administrator $A$ decreases until it becomes negative. Figure~\ref{fig:fig2_exampleTrajectories}B displays the energy output as a function of administration which resembles well the assumed underlying principle of diminishing marginal returns on complexity in \citet{Tainter.2006}. In our first approximation to the theory of collapse, the model society can only react to shocks by choosing the best possible administrator each time the energy requirements are not met. We interpret the formation of an administrative body as an increase in societal complexity. After a period of increased energy output, marginal returns decrease while correspondingly the share of administration rises, caught in an energy-complexity-spiral until the society collapses (here $E = 0$), Fig.~\ref{fig:fig2_exampleTrajectories}A.

Next, we study whether a small adaptation mechanism of the model society enables the model to overcome the fast collapse shown in Fig.~\ref{fig:fig2_exampleTrajectories}A. For this purpose we allow for a random exploration of node states 
enabling the model agents to change their state ($A \longrightarrow L/C$ or vice versa) with a low probability ($p_e$ = 0.00275). For a network of $N=400$ individuals this amounts to an average of 1.1 nodes changing their status at random per time step.

As figure \ref{fig:fig2_exampleTrajectories}C shows, an exploration rate of this low magnitude can already be sufficient to delay the collapse that follows from the event-triggered selection of administrators indicated by a higher survival time of the society, compare Fig.~\ref{fig:fig2_exampleTrajectories}A to Fig.~\ref{fig:fig2_exampleTrajectories}C. However, energy output $E$ is still very low at $p_e = 0.00275$, due to the high degree of administration in the network over sustained periods of time.

We also study the case of larger rates of exploration, i.e, $p_e=0.02$, Fig.~\ref{fig:fig2_exampleTrajectories}D. We find an initially sharp increase in administration, facilitated by the high exploration probability. The trajectory then converges to a stable administrator share slightly above $N_A/N=0.5$, slightly increased by additional effect based promotion of nodes, because energy production per capita is always below the threshold ($\epsilon = 1.0$). After reaching this (meta-)stable state, the model society seems to survive for a long time (at least longer than $t=5000$) while faced with a decreased mean energy production at around $E\approx 0.5$.

It is noteworthy that in each instance of the simulation (Fig.~\ref{fig:fig2_exampleTrajectories}A, C, D) the society reaches an energy production above the energy requirements, which is then however quickly surpassed until the energy level either tends towards zero for zero or low exploration rates in Fig.~\ref{fig:fig2_exampleTrajectories}A,C or converges to a stable regime for large exploration rates in Fig.~\ref{fig:fig2_exampleTrajectories}D. Also note that in the standard "fast-collapse" scenario, the society initially fares better with respect to energy production than under exploration scenarios ($p_e \ge 0$).

\subsection{Deterministic macroscopic approximation of the stochastic micro-model} 

We now derive a macroscopic approximation of the above stochastic micro-model in terms of an ordinary differential equation describing the average time evolution of an aggregate quantity (here the total number $N_A$ of administrators). This time evolution is governed by average transition rates between the three groups. The rates are based on approximations of the probabilities with which individual agents switch between the groups and the assumption that $N$ is large so that the law of large numbers applies.

There are two processes which make the number of administrators change, exploration and targeted recruiting in response to energy demands.
Due to exploration, from the $N_A$ many administrators, on average $p_e N_A$ many leave the administration per unit time, and from the $N - N_A$ many non-administrators, on average $p_e (N - N_A)$ many are hired as administrators per unit time, making a balance of $p_e (N - 2 N_A)$ additional administrators on average per unit time.

For targeted recruiting, we make the simplifying assumption that the density of links inside and between the three groups remains approximately constant and can thus be estimated by the overall link density $\rho$ of the Erd\"os-Renyi network. Then, if $N_A$ is the number of administrators, the probability that a non-administrator is not linked to any of the $N_A$ administrators (and is thus un-coordinated) is approximately
$(1-\rho)^{N_A}$, hence the numbers of un-coordinated and coordinated laborers are approximately
$N_L \approx (N - N_A) (1-\rho)^{N_A}$ and
$N_C \approx (N - N_A) (1 - (1-\rho)^{N_A})$,
hence we have
$E = R e$ with $e = N_L^a + c N_C^b \approx [(N - N_A) (1-\rho)^{N_A}]^a + c [(N - N_A) (1 - (1-\rho)^{N_A})]^b$.

An additional administrator is recruited iff $E/N < \epsilon$, i.e., iff $R < N \epsilon / e$. 
We have Eq.~\ref{eqn:resource} ($R = R_{max}(1 - B)$ with $B \sim \mathrm{Beta}(\alpha, \beta)$), hence the condition $R < N \epsilon / e$ is equivalent to $1 - B < N \epsilon / e R_{max} = N^a \epsilon / e$. As $1 - B \sim \mathrm{Beta}(\beta, \alpha)$.
Hence the probability of $E/N < \epsilon$ is given by the cumulative probability function of the Beta distribution, $P = F(N^a \epsilon / e, \beta, \alpha)$.

The expected number of administrators hired additionally due to shocks per time unit is then also equal to this probability $P$. 

In all, we get the approximation

\begin{align}\label{eqn:approx}
    \frac{dN_A}{dt} &\approx p_e(N - 2 N_A) + F\left(\frac{N^a \epsilon}{[(N - N_A) (1 - \rho)^{N_A}]^a + c [(N - N_A) (1 - (1 - \rho)^{N_A})]^b}, \beta, \alpha\right), \\
    N_C &\approx (N - N_A) (1 - (1 - \rho)^{N_A}), \\
    N_L &\approx (N - N_A) (1 - \rho)^{N_A}
\end{align}
as long as $N_A<N$.

\begin{figure}[!t]
    \centering
    \includegraphics[width = \linewidth]{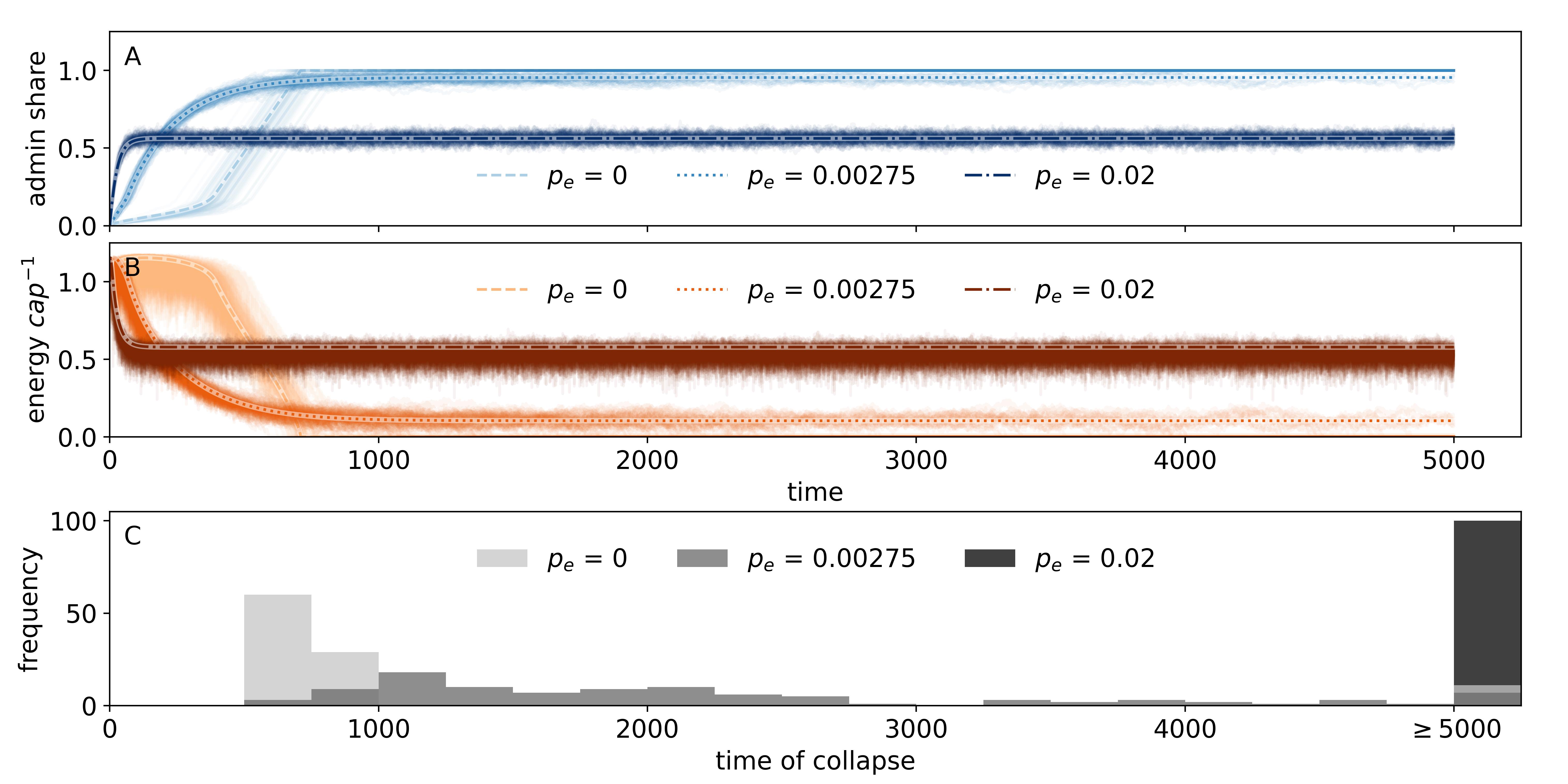}
    \caption{Analytic approximation of the model dynamics with the share of administration $N_A/N$ (A) and corresponding energy production per capita (B). Transparent solid lines show the results of an ensemble of 100 simulations of the microscopic network model for different values of $p_e$. Dashed lines display the respective analytical approximation. (C) Histograms of collapse frequencies of the model, i.e., frequency of times at which energy production approaches zeros, $E=0$. Bars located at time $t \geq 5000$ show right-censored cases of the simulation.}
    \label{fig:fig3_macro_approx}
\end{figure}

Simulation results are well predicted by the approximation, Fig.~\ref{fig:fig3_macro_approx}A,B. Particularly at higher values of $p_e$, the approximation yields excellent results both for the energy production and the administration share. Only at exploration probability of $p_e=0$ (dashed line and light colors in Fig.~\ref{fig:fig3_macro_approx}), the macroscopic approximation slightly overestimates the velocity of collapse. This can be explained by one simplifying assumption of the macroscopic approximation, namely the network degree or local link density to be homogeneous, i.e., approach $\rho$, across the entire network. In contrast, due to the links being put at random when setting up the network, this distribution in a node's number of connections becomes heterogeneous with some nodes showing an above-average number of connections. Because those nodes are preferred over those with few connections in the process of administration selection, initially selected administrators will result in an above average energy return. This produces the prolonged slow-growth phase of the administration in the numerical simulation. In contrast, the macroscopic approximation assumes that at a given point in time any new administrator would have the same average effect. 

To illustrate the probability of collapse under different exploration assumptions, histograms of survival times from the stochastic model are computed, Fig~\ref{fig:fig3_macro_approx}C. At high rates of social mobility $p_e$ the model converges to a stable state where the society persists for infinite times due to its ability absorb shocks. For lower $p_e \approx 0$ we also find isolated cases of infinite survival even though the society mostly collapses comparatively early, Fig.~\ref{fig:fig3_macro_approx}C. Such observations can be explained by a fragmentation of the network into smaller worker networks that are mutually disconnected. The scattered survival times along the time axis for intermediate exploration rates illustrate stochastic survival of the model society due to a random sequences of shocks. 

\subsection{Influence of model parameters on survival time} 

In the third step of our analysis we use the approximation proposed in the previous section to estimate model outcomes for a broad range of parameter values $\rho$, $c$ and $p_e$. Note, that the additional parameters of resource availability ($\alpha$, $\beta$) and threshold ($\epsilon$) have a major influence on the outcome of the model as well. Specifically, a high probability of low resource availability leads to much shorter survival times and vice versa while a low threshold $\epsilon$ to appoint further administrators considerably increases survival times. However, since these effects did not reveal any additionally remarkable results (not shown), we focus our attention on the major drivers of survival time and energy output, i.e, the exploration probability $p_e$, link density $\rho$ and output elasticity of labor $a$. 

\begin{figure}[!t]
    \centering
    \includegraphics[width = \linewidth]{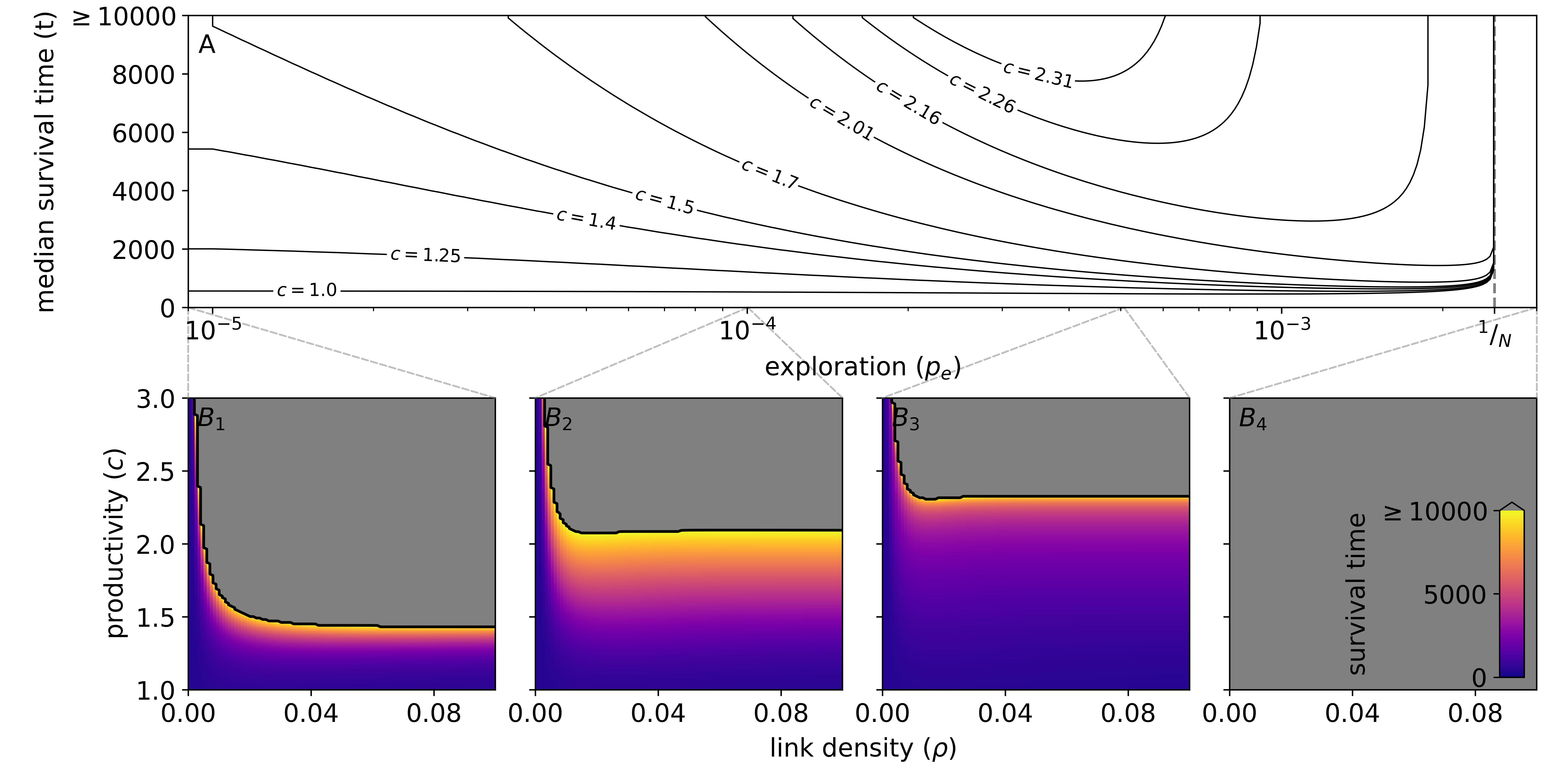}
    \caption{Survival time analysis of the parameters exploration probability ($p_e$), link density ($\rho$) and productivity ($c$).
    (A) The effect of $p_e$ between zero and $\frac{1}{N}$ for some select values of $c$ on median survival time over all computed values of $\rho$. The vertical dashed line defines the exploration threshold of $\frac{1}{N}$.
    Lower panels ($B_{1-4}$) display the relation between $c$ and $\rho$ for different exploration probabilities. Grey areas in the lower panels indicate a survival time $\geq 10000$, potentially going to infinity.}
    \label{fig:f4_parameters}
\end{figure}

Panels \ref{fig:f4_parameters}$\text{B}_{1-4}$ show the survival time as a function of link density $\rho$ and labor productivity $c$. Figure~\ref{fig:f4_parameters}$\text{B}_1$ displays a parameter analysis for the case of no exploration, $p_e=0$ and can hence be interpreted as showing minimum requirements of a successful administration. The most important requirement being a sufficiently large connectivity of the social network (indicated by the increasingly grey area with increasing $\rho$). Also, the extreme case of $\rho = 0.0$, i.e, a network with no connections, must be noted since in this case nodes cannot be converted  at all to administrators and thus the model remains in its initial state. As $p_e$ approaches $\frac{1}{N}$  (Fig.~\ref{fig:f4_parameters}$\text{B}_{2-3}$) the relevance of $\rho$ diminishes and the survival time becomes mainly a function of the productivity $c$. 

\begin{figure}[!t]
    \centering
    \includegraphics[width = \linewidth]{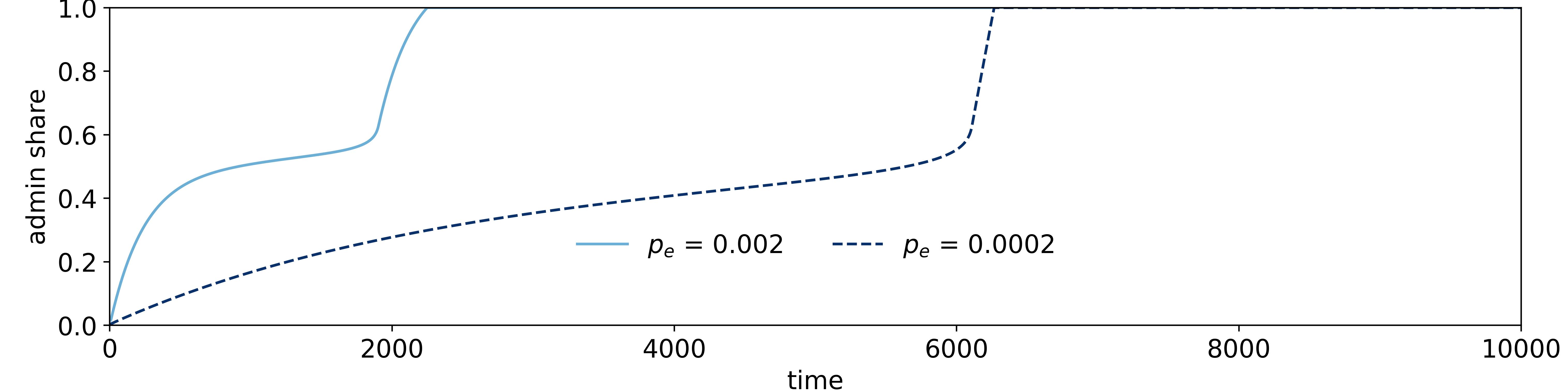}
    \caption{Trajectories of administration share for $c = 2.1$ and $\rho = 0.05$ for two choices of $p_e$ below the critical value $p_e=\frac{1}{N}$ above which collapse of the system is mitigated. The trajectories show how the administration share is stabilized by exploration. This mechanism effectively slows down the total change in administration size until a critical administration size is reached at which administration growth enters the event triggered phase.}
    \label{fig:two_phase_admin_growth}
\end{figure}

Figure \ref{fig:f4_parameters}A demonstrates the effect of exploration for selected values of $a$. 
For exploration probabilities $p_e<1/N$, i.e., when less than one individual per time unit switches their type, collapse is likely, which is indicated by reduced median survival times, Fig.~\ref{fig:f4_parameters}A. 
In fact, our macroscopic approximation shows that $N_A=N$ is a stable fixed point of the approximate dynamics for $p_e < 1/N$ since the RHS of Eqn.\ \ref{eqn:approx} remains strictly positive for $N_A\to N$. This means that the labor force will vanish in finite time, thereby also resulting in a finite survival time.
Additionally, in this ``low exploration regime'', a larger coordinated output productivity $c$ tends to increase the median survival time. 
Indeed, the term $F(\cdot)$ in Eqn.\ \ref{eqn:approx} is a decreasing function of $c$. This effect is the more pronounced, the larger $N_A$ is, because for large $N_A$ the energy production of coordinated laborers $C$ is dominant. Hence a larger $c$ slows down the collapse more and more as we approach $N_A=N$.
The effect of a $p_e$ on survival time is more difficult to understand because it is twofold, as can again be seen in Eqn.\ \ref{eqn:approx}. Initially, when still $N_A<N/2$, larger $p_e$ increases $dN_A/dt$ and speeds up the process of recruiting administrators, as seen on the left of Fig.~\ref{fig:fig3_macro_approx}A. As soon as $N_A>N/2$, the reverse happens and a larger $p_e$ slows down that process. For small $p_e$ below some turning point value $p_e^0<1/N$, the first effect dominates, so that survival time decreases. For $p_e$ between $p_e^0$ and $1/N$, however, the second effect dominates, so that survival time increases again.
For some values of $c$, the two effects are clearly distinguishable in different phases of the evolution, as can be seen in the realization depicted in Fig.~\ref{fig:two_phase_admin_growth}, where the system stays a long time close to $N_A=N/2$ before reaching $N_A=A$ and thereby collapses. 
As $p_e$ approaches the critical value of $1/N$, this effect becomes ever stronger and expected survival times approach infinity.

Of course, since the microscopic network model is of stochastic nature, individual survival times still vary between realizations for the same set of parameters. As survival time also depends on other parameters, we display only the central tendency of survival time in Fig.~\ref{fig:fig3_macro_approx}A. Since we terminate our simulations after $t=10000$ time steps, we cannot distinguish higher survival times, which means we cannot use the arithmetic mean of survival times as the measure of central tendency. This is why we display the median instead, which has the additional advantage of being a more robust statistic than the mean.

For larger values of $c$, we see that some realizations survive for very long even though $p_e<1/N$. This is because for large enough $c$, the approximate dynamics Eqn.\ \ref{eqn:approx} have a second stable fixed point despite $N_A=N$. Indeed, for $c\to\infty$, the $F(\cdot)$ term vanishes and $dN_A/dt=0$ for some value close to $N_A=N/2$. In that case, realizations of the stochastic model starting with small $N_A$ are likely to stay close to $N_A=N/2$ for very long (which is thus a ``metastable'' state) before eventually escaping into the basin of attraction of the ``collapse'' fixed point $N_A=N$ due to a large enough shock. 

When $p_e$ crosses the critical value $1/N$, the collapse fixed point $N_A=N$ of the approximate dynamics becomes unstable, so that collapse can only occur due to a sequence of large shocks which becomes increasingly unlikely. The median survival time is then very large.
For large $p_e$, one can also see from Eqn.\ \ref{eqn:approx} which levels of $N_A$ occur most likely over the course of the simulation. Since the $F(\dots)$ term is between 0 and 1, $dN_A/dt=0$ implies $N/2 \le  N_A \le N/2 + 1/2p$. 
In other words, when either $p_e$ or $c$ are large enough, one can expect that after some transient phase, there will be only slightly more administrators than laborers on average for a long time, before eventually a large shock eventually causes the system collapse after all. 

\section{Discussion}

We have constructed an agent-based network model to illustrate the emergent dynamics of the {\em theory of societal collapse} as postulated in~\citet{Tainter.1988}. In our model, problem solving increases complexity and ultimately leads to system collapse by diminishing marginal returns to investment. Problems were represented abstractly by random shocks to energy production, which were countered by adding complexity to the networked social system in the form of increasing administration. We were able to derive a well-performing macroscopic approximation of the proposed model, which provides a simple mathematical description of the theory in form of an ordinary differential equation. Using this approximation, we showed that increased social mobility through random status exploration ($A \longrightarrow L/C$, $A \longleftarrow L/C$) of at least $p_e=1/N$ is sufficient to mitigate collapse at the expense of skipping a phase of high energy returns on labor. A minimum network link density $\rho \approx 0.02$ was essential for long survival times, but had no further effect beyond a threshold. Finally, the positive effect of complexity, expressed as the productivity $c$ of workers connected to administrators, was found to be the dominant factor increasing survival time and also increasing energy returns. 

Similar to \citet{Motesharrei.2014}, we found that an unequal society in which the elite consumes more (in their case) or produces less (in our case) than the commoners is likely to run into collapse. In addition, we found that the absence of an elite (i.e.\ egalitarian society), which is comparable to the random state exploration mechanism in our model, produced more pathways to avoid collapse than a hierarchical (complex) society. In comparison, previously reported simulations of competing pre-modern societies showed that collapse becomes less likely as societies grow more complex and develop intensive agriculture at sufficiently large scale \citep{Guzman.2018}. This finding contrasts with our result for two specific reasons: First, in \citet{Guzman.2018} complexity is a binary state and cannot continue to increase once a society is denoted as being {\em complex}. Second and more importantly, \citet{Guzman.2018} modeled a warrior class comparable to administrators in our model, which is directly coupled to available food (energy). Therefore, complexity in terms of hierarchies cannot exceed the level of energy available and thereby grow to an unsustainable size. A more detailed investigation into the interactions of complexity and economic development in future work would certainly help in understanding the importance of complexity in collapse. \citet{Butzer.2012} points out that collapse itself should be modeled by sophisticated social-ecological models that do not rely on simplifying assumptions. Nevertheless, \citeauthor{Butzer.2012} concludes that the preconditions for breakdown are typically economic or fiscal decline (i.e. diminishing marginal returns). We agree and emphasize that our model of the theory of {\em collapse of complex societies} offers a potential path leading to exactly such preconditions.

Like in \citet{Watson.1983} or \citet{Granovetter.1978}, the objective of our research was not to derive operational guidelines for building resilient societies, but to illustrate the mechanism of collapse due to diminishing marginal returns of an increasingly complex society. We acknowledge that our approach to modeling collapse is highly conceptual, nevertheless our results underline the relevance of understanding the emergence and consequences of complexity in the resilience discourse. 

One major critique of the model is the built-in ratchet effect, where administrators, once recruited, cannot convert back into productive labor force for $p_e = 0$. This results in a model where collapse is hardwired into the system. 
While this assumption is questionable for real-world scenarios, it serves to exemplify the following statement, which we believe encapsulates J. Tainter's theory: \textit{If the energy available to a system is limited, if a system is confronted with problems that occasionally require it to increase its complexity, if these increases in complexity have associated costs, and the system's complexity cannot be reduced; then, given enough time, the system will inevitably collapse.}
The general validity of assuming persistent complexity is supported by ~\citet{Tainter.1988}, who argues that endogenous reduction of complexity as a problem-solving strategy is not observed in historical cases, at least not as a dominant process. Furthermore, the assumed persistence of complexity can be motivated by persistent power structures \citep{Acemoglu.2008}, which come into existence much more easily than they are removed. Finally, the administration depicted in this model can be compared to expanding elite bureaucracies according to \citet{Parkinson.19.11.1955} and \citet{Weber.1922}. It would be interesting to consider more complex dynamics of intra-generational and intergenerational mobility of social status and wealth and their respective changes over time. 

Unlike more economically spirited models, our simple model does not include any form of competition, economic externality, population growth, or resource depletion, which may be alternative or additional drivers of collapse in the real world. 

Finally, we acknowledge that our proposed model represents only one feature of complexity, i.e., the emergence of one additional layer of hierarchy, administration. Of course, modeling efforts exist that consider much more intricate mechanisms, such as human behavior, shared resource systems, interacting societies, etc. \citep{Arneth.2014, Rounsevell.2011, Schill.2019, Schluter.2017, Nitzbon.2017, Motesharrei.2014}. But keeping Occam's razor in mind, we believe that this very simple model (without population growth, additional hierarchical layers, dynamic networks, etc.) is sufficient to make a case for the direct influence of complexity with respect to social-ecological collapse. 

With the availability of databases of collapse like Seashat \citep{Turchin.2020} and recently published empirical survival analysis of states \citep{Scheffer.2023}, the data are available to ground models on real data. The complexity-generating mechanism proposed in this work can provide the mechanistic basis to simulate the expected slowdown of recovery from perturbations \citep{Scheffer.2023}, as the marginal returns on complexity decrease with increasing complexity of the society. The logical next step is to use survival analysis to combine empirical data with mechanistic models by modeling hazards through advanced mechanistic models in order to test specific hypotheses of societal collapse. For this, the presented model needs to overcome its deficiencies and include resource acquisition and population dynamics not as abstract quantities like energy $E$, but physically based quantities. Therefore, our model should be extended in future studies to analyze how complexity relates to collapse in potentially more realistic scenarios, which could be represented by further social stratification or more complex and adaptive network topologies. In this context, complexity costs of social connectivity should be considered. In addition, impacts of innovations and energy substitutes (coal, oil, etc.) on collapse could be studied to identify additional strategies to slow down or avoid collapse~\citep{lenton2016revolutions}. Furthermore, the interrelation between energy and complexity could also be studied in more elaborate World-Earth and Human-Earth system models to study under which circumstances the results obtained in this work could be transferred to less abstract applications. 
For this purpose, appropriate measures of energy, complexity and resilience need to be defined depending on the specific model and research question at hand \citep{Tamberg.2022}.

With regard to the open research topics posed by \citet{Cumming.2017}, such as moving the debate from `whether' to `why' collapse occurred and possible strategies to ``avoid, slow or hasten collapse'', we wish to contribute our findings to the discourse on the drivers of social-ecological resilience and collapse. Growing complexity has recently been identified as a source of increased societal risk, e.g. due to high levels of socio(-ecological) interconnectedness that can lead to cascading failures \citep{Helbing.2013, Rocha.2018}. In this regard complexity is mainly considered as a facilitator of collapse by shaping systems to become more vulnerable. In our research, we show that complexity may be even more directly linked to collapse, as societies may reach a point where the associated costs outweigh the benefits, thereby directly causing their breakdown. This dimension is, to our understanding, an underrepresented view of social-ecological collapse that should be strengthened by future research efforts. 

While increasing complexity has brought vast benefits to modern industrialized societies, our findings also raise the question of how far contemporary societies can increase complexity without creating large-scale risks of collapse due to ever-increasing energy demands. Along those lines, it is of great interest to further investigate whether increasing complexity is always associated with increased risk, or whether certain forms of complexity exist that do not necessarily increase the risk of societal collapse. Such an assessment would allow the development of design principles for resilient infrastructures and social structures that do not facilitate complexity in the understanding of~\citet{Tainter.1988}.


\subsection*{Author Contributions}

Conceptualization, Jobst Heitzig and Jonathan Donges; Data curation, Florian Schunck; Formal analysis, Florian Schunck; Funding acquisition, Jonathan Donges; Investigation, Florian Schunck, M. Wiedermann, Jobst Heitzig and Jonathan Donges; Methodology, M. Wiedermann and Jobst Heitzig; Project administration, Jonathan Donges; Software, Florian Schunck; Supervision, Jobst Heitzig and Jonathan Donges; Validation, Jobst Heitzig; Visualization, Florian Schunck and M. Wiedermann; Writing – original draft, Florian Schunck, M. Wiedermann and Jobst Heitzig; Writing – review \& editing, Florian Schunck, M. Wiedermann, Jobst Heitzig and Jonathan Donges.

\subsection*{Funding}

The work was carried out in the scope of the COPAN collaboration at the Potsdam Institute for Climate Impact Research. M.W.\, J.H.\ and J.F.D.\ were supported by the Leibniz Association (project DominoES) and the European Research Council (ERC) under the European Union's Horizon 2020 Research and Innovation Programme (ERC grant agreement No.\ 743080 ERA). F.S. was supported by the German National Academic Foundation. J.F.D. is grateful for financial support by the German Federal Ministry for Education and Research (BMBF) under grant 01LS2001A (project CHANGES). The authors  gratefully acknowledge the European Regional Development Fund (ERDF), the BMBF and the Land Brandenburg for providing resources on the high-performance computer system at PIK.

\subsection*{Data availability}

The stochastic model as well as the macroscopic approximation that produced the results in this work are openly available in the Github repository \hyperlink{https://github.com/flo-schu/Tainter}{https://github.com/flo-schu/Tainter}. The repository contains all necessary information to install the package and download pre-simulated datasets hosted on the Open Science Foundation (\hyperlink{https://osf.io/u897c/}{https://osf.io/u897c/})

\subsection*{Acknowledgements}

We thank the German National Academic Foundation (Studienstiftung des deutschen Volkes) for facilitating the student college that initiated this project. We are grateful to the comments of three anonymous reviewers from a previous submission, which have significantly advanced this manuscript. For fruitful discussions we sincerely thank Jürgen Kurths, Lea Tamberg, Adrian Lison, Sascha Haupt, Vivian Grudde, Bastian Grudde (née Ott) and other members of the Studienkolleg's working group 4. We thank Wolfgang Lucht for early inspiring discussions that triggered our interest in Tainter's theory. We employed artificial intelligence tools to refine the wording and grammar of our manuscript

\subsection*{The authors declare no conflict of interest.}

\section*{Appendix A: Optimal number of administrators}

From the analytical approximation, we know that the expected energy input $E$ is proportional to

\begin{align}
    E_0 &:= [(N - N_A) (1 - \rho)^{N_A}]^a + c[(N - N_A) (1 - (1 - \rho)^{N_A})]^b,
\end{align}

which has a maximum in $N_A$ where

\begin{align}
    0 = \partial_{N_A} E_0 &=
    - a[(N - N_A) (1 - \rho)^{N_A}]^{a-1} [1 - (N - N_A) \ln(1 - \rho)](1 - \rho)^{N_A} \\
    &+ bc[(N - N_A) (1 - (1 - \rho)^{N_A})]^{b - 1}
    [(1 - \rho)^{N_A} [1 - (N - N_A) \ln(1 - \rho)] - 1].
\end{align}

E.g., for $N=400$, $\rho=0.02$, $a=b=0.75$ and $c = 1.05$ the optimal number of administrators is $N_A\approx 22$ or $\approx 6\%$, leading to an $E\approx 461$ or $E/N\approx 1.15$.

\bibliography{references} 

\end{document}